\documentclass[english,aps,preprint]{revtex4-1}
\usepackage[T1]{fontenc}
\usepackage[latin9]{inputenc}
\usepackage{units}
\usepackage{amsmath}
\usepackage{graphicx}

\begin{document}
\title{Spin entanglement signatures of proton from a light-front Hamiltonian}

\author{Chen Qian}
\email{qianchen@baqis.ac.cn}
\affiliation{Beijing Academy of Quantum Information Sciences, Beijing 100193,
China}

\author{Siqi Xu}
\email{xsq234@163.com}
\affiliation{Institute of Modern Physics, Chinese Academy of Sciences,
Lanzhou 730000, China}
\affiliation{Department of Physics and Astronomy, Iowa State University, Ames, Iowa 50011, USA}

\author{Yang-Guang Yang}
\email{yyg@impcas.ac.cn}
\affiliation{Institute of Modern Physics, Chinese Academy of Sciences,
Lanzhou 730000, China}
\affiliation{School of Nuclear Science and Technology, University of Chinese
Academy of Sciences, Beijing 100049, China}

\author{Xingbo Zhao}
\email{xbzhao@impcas.ac.cn}
\affiliation{Institute of Modern Physics, Chinese Academy of Sciences,
Lanzhou 730000, China}
\affiliation{School of Nuclear Science and Technology, University of Chinese
Academy of Sciences, Beijing 100049, China}

\begin{abstract}
Quantum entanglement provides a quantitative probe of the internal structure of hadrons and offers a sensitive means to study the quantum correlation in the hadron wave functions. For baryons, the spin state of the three valence quarks forms a tripartite qubit system, whose entanglement structure can be characterized by the four classes of three-qubit states. In this work, we compare the proton spin entanglement obtained from Basis Light-Front Quantization (BLFQ) with that from a quark-diquark model. By analyzing both bipartite and tripartite entanglement, we find that the quark-diquark model yields a substantially more entangled spin state than the BLFQ wave function in the valence Fock sector. This difference mainly originates from the larger W-type and Bell-type entanglement in the quark-diquark model. Within BLFQ, larger stronger coupling constant and smaller quark mass drive the spin correlation among the valence quarks towards an effective quark-diquark configuration with an active $d$ quark and a correlated $uu$ pair.
\end{abstract}
\maketitle

\section{Introduction\label{sec:Introduction}}

Hadrons are relativistic quantum many-body systems of quarks and gluons formed by the strong interaction. Their structure can be described by the fundamental theory of strong interaction, Quantum Chromodynamics (QCD). Hadron structure is usually characterized by observables representing the probability distributions of quarks and gluons, such as the form factors~\cite{Ma:2002ir,Zhang:2016qqg,Adhikari:2016idg,Adhikari:2018umb,Meng:2024pkc}, leading-twist generalized parton distributions~\cite{deTeramond:2018ecg,Liu:2022fvl,Lin:2023ezw} and transverse momentum dependent parton distributions (TMDs)~\cite{Hu:2020arv,Hu:2022ctr}. While these quantities provide a detailed description of hadron structure on the probability level, they lack the information on how quantum correlation is organized and distributed among constituents inside the hadron. 

Recently quantum-information-theoretic quantities have been proposed as powerful new probes of hadron structure. For instance, entanglement entropy has been applied to describe the quantum correlation of partons in deep inelastic scattering~\cite{Kharzeev:2017qzs,Tu:2019ouv,Gotsman:2020bjc,Kharzeev:2021yyf,Zhang:2021hra} and hadronization processes~\cite{Datta:2024hpn}. In these processes, quantum-information studies offer an access to structural features of hadrons which cannot be captured by the conventional observables. On the theory side, the hadron wave functions from different models exhibit distinct patterns of quantum correlation. Taking the proton as an example, in the quark-diquark model, the proton's wave function is formulated in terms of an active quark and a strongly correlated pair of quarks named diquark~\cite{BRODSKY20151,deTeramond:2008ht,Bacchetta:2008af,Chakrabarti:2015ama,Maji:2016yqo}, whereas the wave function from the Basis Light-Front Quantization (BLFQ) approach treats each quark as an independent degree of freedom~\cite{Vary:2009gt,Mondal:2019jdg,Xu:2023nqv,Xu:2021wwj,Xu:2024sjt}. Consequently, the wave functions from these two approaches show distinct entanglement patterns. To systematically characterize and compare the patterns of quantum correlation across different theoretical models, quantum-information-theoretic quantities such as the entanglement entropy can serve as a new diagnostic tool~\cite{Qian:2024fqf,Dumitru:2022tud,Dumitru:2023qee,Dumitru:2025bib,Zhang:2025ean}.

Following our previous work~\cite{Qian:2024fqf}, in this work we study the structure of quantum correlation between the valence quarks inside a proton by concentrating on the spin entanglement in the proton wave function. As an example, we compare the wave functions from two approaches: a quark-diquark model motivated by light-front holographic QCD (LFHQCD) and the Basis Light-Front Quantization (BLFQ) approach. We find that the protons constructed from these different models exhibit distinct entanglement structure among partons. To analyze the proton wave function with a quark-diquark structure, we resolve the diquark as a correlated two-quark cluster to enable a direct comparison. On the BLFQ side, for the convenience of comparison, we work with the proton wave function truncated to the leading $\left|qqq\right\rangle$ Fock sector. After that, we investigate bipartite and tripartite entanglement among the valence quarks based on the proton wave functions from these two approaches. We use the entanglement entropy to quantify the bipartite entanglement between the active quark and the remaining system, and the $\pi$-tangle as well as the $F_{123}$ triangle to quantify the genuine tripartite entanglement among the three quarks. The results suggest that the entanglement between the quarks in the quark-diquark wave function is much stronger than that from BLFQ, and this feature is robust with respect to different choices of input parameters within the BLFQ framework. To better understand these results, we invoke the tripartite state classification~\cite{Dur:2000zz} and show that the stronger entanglement in the quark-diquark wave function originates from the exchange symmetry in the Bell-entangled diquark states, which induces substantial W-type entanglement. Furthermore, we note that with the increase of the strong coupling constant in the input Hamiltonian, the BLFQ wave function develops an effective quark-diquark structure, where the $d$ quark acts as the active quark and the $uu$ quark pair forms the diquark cluster.

This paper is organized as follows. We briefly review the concepts of bipartite and tripartite qubit entanglement measures in Sec.~\ref{sec:methodology}. Subsequently, we review the quark-diquark model based on LFHQCD and derive its spin states in Sec.~\ref{sec:quark-diquark-or-not}. Next, we review the basics of the BLFQ approach and compare the resulting proton wave function with that from the quark-diquark model in terms of the bipartite and tripartite entanglement in Sec.~\ref{sec:Compare-spin-state-BLFQ}. After that, we examine the sensitivity of  the spin entanglement patterns in the BLFQ wave functions with respect to different choices of input parameters in the light-front Hamiltonian in BLFQ in Sec.~\ref{sec:Spin-ent-para}. Finally, we summarize and give an outlook in Sec.~\ref{sec:Conclusions-and-outlook}.

\section{Basic concepts of entanglement measures\label{sec:methodology}}

This section begins with a brief overview of the bipartite and tripartite entanglement quantities employed in this work. For a system in a pure state, the entanglement entropy provides the most direct indicator of bipartite entanglement, which has been introduced to the proton wave functions in Ref.~\cite{Qian:2024fqf}. That is, for two complementary parts $\mathrm{A}$ and $\mathrm{B}$ of a pure state in one $N$-partite system $\left|\psi\right\rangle$, the entanglement between them can be characterized by Von-Neumann entanglement entropy:
\begin{equation}
S=-\mathrm{Tr}\left(\rho_{\mathrm{A}}\mathrm{\log_{2}\rho_{A}}\right)=-\mathrm{Tr}\left(\rho_{\mathrm{B}}\mathrm{\log_{2}\rho_{B}}\right),\;\label{eq:ent-entropy}
\end{equation}
where $\rho_{\mathrm{A}}=\mathrm{Tr_{B}}\rho_{\mathrm{AB}}$ and $\rho_{\mathrm{B}}=\mathrm{Tr_{A}}\rho_{\mathrm{AB}}$ with $\rho_{AB}=|\psi\rangle\langle\psi|$ are called reduced density matrices of subsystems $\mathrm{A}$ and $\mathrm{B}$, respectively. For a given $N$-partite system, the maximum value of its entanglement entropy is $S=\mathrm{min}\{N_{\mathrm{A}}, N_{\mathrm{B}}\}$, representing the maximal entanglement between subsystems $\mathrm{A}$ and $\mathrm{B}$, where $N_A$ and $N_B$ denote the dimension of $\mathrm{A}$ and $\mathrm{B}$, respectively.

Compared to the bipartite entanglement, the tripartite entanglement between the three valence quarks can provide additional structural information in a baryon wave function in the valence Fock sector. Here, we briefly review the classification of tripartite qubit states, and introduce two quantities characterizing the genuine tripartite entanglement for both pure states and mixed states~\cite{Dur:2000zz,Miyake:2003tee}, namely the triangle measure $F_{123}$~\cite{Xie:2021hsy,Horodecki:2025tpn,Sakurai2024} and the $\pi-$tangle~\cite{Ou:2007ysv,Osterloh:2007cwe,Kim:2008ava}.

In recent years, the notion of genuine multipartite entanglement has become widely adopted in the study of composite quantum systems. For three qubits, D{\"u}r, Vidal, and Cirac~\cite{Dur:2000zz} showed that all states fall into four distinct classes: product states, biseparable states, the Greenberger-Horne-Zeilinger (GHZ) class, and the W class. In the first two classes at least one qubit is unentangled with the remainder, whereas the three qubits in the GHZ and W classes are genuinely entangled. Encoding the spin state $\left|\uparrow\right\rangle =\left|1\right\rangle $ and $\left|\downarrow\right\rangle=\left|0\right\rangle$, the GHZ and W states are written as,
\begin{eqnarray}
\left|\mathrm{GHZ}\right\rangle  & = & \frac{1}{\sqrt{2}}\left(\left|000\right\rangle \pm\left|111\right\rangle \right),\;\label{eq:GHZ-state}\\
\left|\mathrm{W}\right\rangle ^{+} & = & \frac{1}{\sqrt{3}}\left(\left|011\right\rangle +\left|101\right\rangle +\left|110\right\rangle \right),\;\label{eq:W-plus}\\
\left|\mathrm{W}\right\rangle ^{-} & = & \frac{1}{\sqrt{3}}\left(\left|100\right\rangle +\left|010\right\rangle +\left|001\right\rangle \right),\;\label{eq:W-minis}
\end{eqnarray}
where the components $\left\{ \left|001\right\rangle ,\left|010\right\rangle ,\left|100\right\rangle \right\}$ in $\left|\mathrm{W}\right\rangle ^{-}$ are called $1-$excitation Dicke states and the components $\left\{ \left|011\right\rangle ,\left|101\right\rangle ,\left|110\right\rangle \right\}$ in $\left|\mathrm{W}\right\rangle ^{+}$ are called $2-$excitation Dicke states~\cite{Dicke:1954zz,Barnea:2014kvt}. To characterize the entanglement structure in the GHZ and W states, without loss of generality, we trace out the third qubit from their density matrices, and the remaining part of GHZ and W is,
\begin{eqnarray*}
\rho_{12}^{\mathrm{GHZ}} & = & \frac{1}{2}\left(\left|00\right\rangle \left\langle 00\right|+\left|11\right\rangle \left\langle 11\right|\right),\;\\
\rho_{12}^{\mathrm{W^{-}}} & = & \frac{1}{3}\left(\left|00\right\rangle \left\langle 00\right|+\left|01\right\rangle \left\langle 01\right|+\left|10\right\rangle \left\langle 10\right|+\left|01\right\rangle \left\langle 10\right|+\left|10\right\rangle \left\langle 01\right|\right).
\end{eqnarray*}
Here we take $|\mathrm{W}\rangle^{-}$ as an example for the W states. One note that $\rho_{12}^{\mathrm{GHZ}}$ is a maximally mixed state, which indicates there is no entanglement between qubit $1$ and qubit $2$. In contrast, the elements in $\rho_{12}^{\mathrm{W^{-}}}$ are similar with the Bell state $\frac{1}{\sqrt{2}}\left(\left|01\right\rangle+\left|10\right\rangle\right)$, but with an additional nonzero diagonal element $\left|00\right\rangle \left\langle 00\right|$. Therefore, the entanglement in GHZ states is totally genuine tripartite entanglement while the entanglement in W states is both bipartite and tripartite.

One of the commonly used quantities to describe the genuine tripartite entanglement is called triangle measure, which corresponds to the area of the concurrence triangle. This quantity was proposed by Xie and Eberly in Ref.~\cite{Xie:2021hsy}, which satisfies all the eligibility criteria for characterizing tripartite entanglement: vanishing for all product and biseparable states, being positive for all non-biseparable states, and not increasing under Local Operations and Classical Communication (LOCC). The quantity $F_{123}$ has the form of
\begin{equation}
F_{123}=\left[\frac{16}{3}\,Q\,(Q-\mathcal{C}^2_{1(23)})(Q-\mathcal{C}^2_{2(13)})(Q-\mathcal{C}^2_{3(12)})\right]^{\tfrac{1}{4}},\;\label{eq:F123}
\end{equation}
where $Q=\frac{1}{2}\left(\mathcal{C}_{1(23)}^{2}+\mathcal{C}_{2(13)}^{2}+\mathcal{C}_{3(12)}^{2}\right)$, and $\mathcal{C}_{i\left(kj\right)}$ is the concurrence between $i$ and the composite subsystem ($kj$). When the tripartite state is a pure state $\left|\Psi\right\rangle _{123}$, the $\mathcal{C}_{i\left(kj\right)}$ is computed as 
\begin{equation}
\mathcal{C}_{i\left(kj\right)}=\sqrt{2\left(1-\mathrm{Tr}\rho_{kj}^{2}\right)},\;\label{eq:concurrence}
\end{equation}
where $\rho_{kj}$ is the reduced density $\rho_{kj}=\mathrm{Tr}_{i}\left(\left|\Psi\right\rangle _{123}\left\langle \Psi\right|\right)$, with $i=1,2,3$ and $j,k=\left\{ 2,3\right\} ,\left\{ 1,3\right\} ,\left\{ 1,2\right\}$. In this definition, $F_{123}$ takes values between $0$ and $1$. Additionally, the concurrence for the bipartite reduced density matrix $\rho_{ij}$ is defined as
\begin{equation}
\mathcal{C}_{ij}=\mathrm{max}\left(0,\lambda_1-\lambda_2-\lambda_3-\lambda_4\right),\;\label{eq:biconcurrence}
\end{equation}
where $\lambda_{i}$ are the eigenvalues of the Hermitian matrix $R=\sqrt{\sqrt{\rho_{ij}}\tilde{\rho_{ij}}\sqrt{\rho_{ij}}}$ with $\tilde{\rho_{ij}}=(\sigma_y\otimes\sigma_y)\rho_{ij}^{*}(\sigma_y\otimes\sigma_y)$ in the decreasing order, and $\rho_{ij}^{*}$ denotes the complex conjugate of $\rho_{ij}$. For the GHZ states, the bipartite concurrences are zero $\mathcal{C}_{12}=\mathcal{C}_{13}=\mathcal{C}_{23}=0$ but $F_{123}=1$. For the W states, the bipartite concurrences are all $\sim 0.667$ and $F_{123}\sim 0.889$.

In addition to the triangle measure $F_{123}$, we introduce another quantity to characterize the tripartite entanglement in the W and GHZ states, namely the $\pi-$tangle. The tangle is a quantity formulated by the difference between the tripartite concurrence and the bipartite concurrence in the form of $\tau_{123}=\mathcal{C}_{1\left(23\right)}^{2}-\mathcal{C}_{12}^{2}-\mathcal{C}_{13}^{2}$. For the W state, $\tau_{123}=0$ and for the GHZ state, $\tau_{123}=1$~\cite{Coffman:1999jd}. However, compared to the triangle measure, the tangle has some shortcomings~\cite{Lohmayer:2006wms}. One of them is that the function $\tau_{123}$ violates monotonicity for a superposition of W state and GHZ state with probability $p$ and 1$-p$ respectively; another is that it cannot distinguish W-type entanglement and biseparable entanglement because for both we have $\tau_{123}=0$. To overcome these drawbacks, $\pi-$tangle was introduced in terms of the concept of negativity~\cite{Ou:2007ysv}. For an arbitrary tripartite state $\rho_{123}$, the negativity is defined as:
\begin{equation}
\mathcal{N}_{i|kj}\left(\rho_{123}\right)=\sum_{\lambda_{i}<0}\left|\lambda_{i}\right|,\;\label{eq:neg}
\end{equation}
where $\left\{ \lambda_{i}\right\} $ are eigenvalues of $\rho_{123}^{T_{i}}$, and $T_{i}$ is the partial transpose with respect to the subsystem $i$. For the bipartite reduced density matrix $\rho_{ij}$, the negativity takes a similar form of
\begin{equation}
\mathcal{N}_{ij}\left(\rho_{ij}\right)=\sum_{\lambda_{j}<0}\left|\lambda_{j}\right|,\;\label{eq:neg_bi}
\end{equation}
where $\left\{ \lambda_{j}\right\} $ are eigenvalues of $\rho_{ij}^{T_{j}}$, and $T_{j}$ is the partial transpose with respect to the subsystem $j$. In terms of the negativity, one can define the $\pi$-tangle as
\begin{equation}
\pi_{i}=\mathcal{N}_{i|kj}^{2}\left(\rho_{123}\right)-\mathcal{N}_{ij}^{2}\left(\rho_{ij}\right)-\mathcal{N}_{ik}^{2}\left(\rho_{ik}\right),\;\label{eq:pi-tangle}
\end{equation}
where $i=1,2,3$ and $j,k=\left\{ 2,3\right\} ,\left\{ 1,3\right\} ,\left\{ 1,2\right\}$. As the final step, we calculate the value averaged over $\pi_i$ as,
\begin{equation}
\pi=\frac{1}{3}\left(\pi_{1}+\pi_{2}+\pi_{3}\right).\;\label{eq:pi-value}
\end{equation}
The value of $\pi-$tangle varies for states with different structure of tripartite entanglement. For product states or biseparable states, $\pi=0$; for W states, $\pi\sim0.1373$; and for GHZ states, $\pi=0.25$.

\section{Spin correlation in the quark-diquark structure\label{sec:quark-diquark-or-not}}

The diquark is a pair of correlated quarks behaving like a single constituent inside a hadron. The notion of the diquark has long been adopted to simplify the description of baryon from the complicated three-body problem to a more tractable two-body problem. Various quark-diquark models have been proposed and they have made remarkable successes in explaining baryon spectroscopy and electromagnetic properties~\cite{Anselmino:1992vg}. Since the light-front quantization offers a powerful framework for describing relativistic bound states, such as the hadrons in QCD~\cite{Brodsky:1997de}, in this work we consider a light-front quark-diquark model~\cite{Bacchetta:2008af,Maji:2016yqo} to study its spin entanglement structure.

In this model, the wave function of a spin-$1/2$ proton with momentum $P$ takes the form of
\begin{equation}
\left|P,J_{z}=\pm\frac{1}{2}\right\rangle =C_{S}\left|uS^{0}\right\rangle ^{\pm}+C_{V}\left|uA^{0}\right\rangle ^{\pm}+C_{VV}\left|dA^{1}\right\rangle ^{\pm},\;\label{eq:proton-di}
\end{equation}
where $\left|uS^{0}\right\rangle $ is the component of scalar spin with scalar isospin, $\left|uA^{0}\right\rangle$ is the component of vector spin with scalar isospin, and $\left|dA^{1}\right\rangle$ is the component of vector spin with vector isospin. $C_S$, $C_V$ and $C_{VV}$ are the coefficients of these three components. They are determined phenomenologically from fitting the proton's electromagnetic form factors~\cite{Maji:2016yqo}. In terms of the quark-diquark degrees of freedom, the $\left|qS\right\rangle ^{\pm}$ and $\left|qA\right\rangle ^{\pm}$ components can be written as,
\begin{equation}
\left|qS\right\rangle ^{\pm}=\int\frac{dx\,d^{2}\mathbf{p}_{\perp}}{2(2\pi)^{3}\sqrt{x(1-x)}}\Big[\psi_{+}^{\pm(q)}(x,\mathbf{p}_{\perp})\left|+\frac{1}{2},0;xP^{+},\mathbf{p_{\perp}}\right\rangle +\psi_{-}^{\pm(q)}(x,\mathbf{p}_{\perp})\left|-\frac{1}{2},0;xP^{+},\mathbf{p_{\perp}}\right\rangle \Big],\;\label{eq:qS-di}
\end{equation}
and
\begin{eqnarray}
\left|qA\right\rangle ^{\pm}=\int\frac{dx\,d^{2}\mathbf{p}_{\perp}}{2(2\pi)^{3}\sqrt{x(1-x)}}\Big[ & \psi_{++}^{\pm(q)}(x,\mathbf{p}_{\perp})\left|+\frac{1}{2},+1;xP^{+},\mathbf{p}_{\perp}\right\rangle +\psi_{-+}^{\pm(q)}(x,\mathbf{p}_{\perp})\left|-\frac{1}{2},+1;xP^{+},\mathbf{p}_{\perp}\right\rangle \nonumber \\
 & +\psi_{+0}^{\pm(q)}(x,\mathbf{p}_{\perp})\left|+\frac{1}{2},0;xP^{+},\mathbf{p}_{\perp}\right\rangle +\psi_{-0}^{\pm(q)}(x,\mathbf{p}_{\perp})\left|-\frac{1}{2},0;xP^{+},\mathbf{p}_{\perp}\right\rangle \nonumber \\
 & +\psi_{+-}^{\pm(q)}(x,\mathbf{p}_{\perp})\left|+\frac{1}{2},-1;xP^{+},\mathbf{p}_{\perp}\right\rangle +\psi_{--}^{\pm(q)}(x,\mathbf{p}_{\perp})\left|-\frac{1}{2},-1;xP^{+},\mathbf{p}_{\perp}\right\rangle \Big],\nonumber \\
 \label{eq:qA-di}
\end{eqnarray}
where $x$ is the longitudinal momentum fraction carried by the quark and $p_\perp$ is the relative transverse momentum between the quark and diquark. The amplitudes of the helicity components $\psi_{\pm}^{\pm(q)}(x,\mathbf{p}_{\perp})$, $\psi_{\pm0}^{\pm(q)}(x,\mathbf{p}_{\perp})$ and $\psi_{\pm\pm}^{\pm(q)}(x,\mathbf{p}_{\perp})$ have been calculated in Ref.~\cite{Maji:2016yqo}. In order to study the entanglement between individual quarks and make comparison with the BLFQ wave function, we expand the diquark into a correlated two-quark cluster. However, we need to keep in mind that treating the diquark as a correlated two-quark cluster, e.g. $u[ud-du]$ for the scalar diquark, introduces additional internal degrees of freedom and may, as a result, increase the overall level of entanglement in the proton's wave function.

For the convenience of discussion, in what follows we take the proton spin-up state $|P, +1/2\rangle$ as an example and encode the spin-up state $\left|+\tfrac{1}{2}\right\rangle$ as $\left|1\right\rangle$ and the spin-down state $\left|-\tfrac{1}{2}\right\rangle$ as $\left|0\right\rangle$. In this work, since our focus is on the quantum entanglement on the spin degrees of freedom, we integrate out the momentum degrees of freedom and retain only the spin degrees of freedom in the light-front wave function of the proton. Following this procedure, after resolving the internal spin structure of the diquarks in terms of their two constituent quarks and integrating over the momentum degrees of freedom, we write down the spin isoscalar-scalar component as,
\begin{eqnarray}
\left|uS^0\right\rangle ^{+} & = & \frac{1}{2}S_{+}^{u_{1}\left[d_{2}u_{3}\right]}\left(\left|1_{1}0_{2}1_{3}\right\rangle -\left|1_{1}1_{2}0_{3}\right\rangle \right)+\frac{1}{2}S_{-}^{u_{1}\left[d_{2}u_{3}\right]}\left(\left|0_{1}0_{2}1_{3}\right\rangle -\left|0_{1}1_{2}0_{3}\right\rangle \right)\nonumber \\
 & + & \frac{1}{2}S_{+}^{\left[u_{1}d_{2}\right]u_{3}}\left(-\left|0_{1}1_{2}1_{3}\right\rangle +\left|1_{1}0_{2}1_{3}\right\rangle \right)+\frac{1}{2}S_{-}^{\left[u_{1}d_{2}\right]u_{3}}\left(-\left|0_{1}1_{2}0_{3}\right\rangle +\left|1_{1}0_{2}0_{3}\right\rangle \right),\;\label{eq:spin-qS}
\end{eqnarray}
where we have expanded the scalar diquark to a spin singlet state $(|01\rangle-|10\rangle)/\sqrt{2}$, which introduces a factor of $1/\sqrt{2}$; and we have taken into account the anti-symmetry under the exchange of the two quarks in the diquark, which introduces another factor of $1/\sqrt{2}$. $S_{+}^{\cdots}$ and $S_{-}^{\cdots}$ are normalized amplitudes evaluated from Eq.~$\left(\ref{eq:qS-di}\right)$, such that the proton state is normalized as $\langle P, J_z=1/2|P, J_z=1/2\rangle=1$, see Eq.~\eqref{eq:spin-ent} below. Here for simplicity we ignore the possible phases in these amplitudes and treat $S^{\cdots}_+$ and $S^{\cdots}_-$ as positive real numbers. Eq.~\eqref{eq:spin-qS} serves as our model wave function for describing the spin degrees of freedom in the proton, which is referred to as the ``spin state'' in the following part of this work. Here we notice that In the quark-diquark model, the spin state of the proton can be considered as a superposition of an active quark times the entangled Bell state of the remaining two quarks, that is, the diquark~\cite{Chakrabarti:2015ama,Maji:2016yqo,Forkel:2008un}. Because in the $|uS^0\rangle$ component the isospin of the isoscalar diquark cluster is $0$ as well, the flavor of the active quark can only be $u$. Since the quark-diquark wave function is symmetric with respect to $u_1$ and $u_3$, we can choose either $u_1$ or $u_3$ as the active one without loss of generality, that is, $S_{\pm}^{u_{1}[d_{2}u_{3}]} = S_{\pm}^{[u_{1}d_{2}]u_{3}} = S_{\pm}^{u}$.

Similarly, we can apply the same expansion to the isoscalar-vector diquark component $\left|uA^0\right\rangle ^{\pm}$ and the isovector-vector diquark component $\left|dA^1\right\rangle ^{\pm}$. Notice that with the total $J_{z}=+\frac{1}{2}$, the contribution of the $\left|\pm\frac{1}{2},-1;xP^{+},\mathbf{p}_{\perp}\right\rangle$ component is zero in this model~\cite{Maji:2016yqo}. Explicitly, we expand the component $\left|uA^0\right\rangle^{+}$ by permuting $\left|u_{1}\left[d_{2}u_{3}\right]\right\rangle$ and $\left|\left[u_{1}d_{2}\right]u_{3}\right\rangle$ as
\begin{eqnarray}
\left|uA^0\right\rangle ^{+} & = & \frac{1}{\sqrt{2}}A_{++}^{u_{1}\left[d_{2}u_{3}\right]}\left|1_{1}1_{2}1_{3}\right\rangle +\frac{1}{\sqrt{2}}A_{-+}^{u_{1}\left[d_{2}u_{3}\right]}\left|0_{1}1_{2}1_{3}\right\rangle \nonumber \\
 & + & \frac{1}{2}A_{+0}^{u_{1}\left[d_{2}u_{3}\right]}\left(\left|1_{1}0_{2}1_{3}\right\rangle +\left|1_{1}1_{2}0_{3}\right\rangle \right)+\frac{1}{2}A_{-0}^{u_{1}\left[d_{2}u_{3}\right]}\left(\left|0_{1}0_{2}1_{3}\right\rangle +\left|0_{1}1_{2}0_{3}\right\rangle \right)\nonumber \\
 & + & \frac{1}{\sqrt{2}}A_{++}^{\left[u_{1}d_{2}\right]u_{3}}\left|1_{1}1_{2}1_{3}\right\rangle +\frac{1}{\sqrt{2}}A_{-+}^{\left[u_{1}d_{2}\right]u_{3}}\left|1_{1}1_{2}0_{3}\right\rangle \nonumber \\
 & + & \frac{1}{2}A_{+0}^{\left[u_{1}d_{2}\right]u_{3}}\left(\left|0_{1}1_{2}1_{3}\right\rangle +\left|1_{1}0_{2}1_{3}\right\rangle \right)+\frac{1}{2}A_{-0}^{\left[u_{1}d_{2}\right]u_{3}}\left(\left|0_{1}1_{2}0_{3}\right\rangle +\left|1_{1}0_{2}0_{3}\right\rangle \right);\label{eq:spin-uA}
 \end{eqnarray}
 and $\left|dA^1\right\rangle ^{+}$ by permuting $\left|\left[u_{1}\right]d_{2}\left[u_{3}\right]\right\rangle$ as
 \begin{eqnarray}
 \left|dA^1\right\rangle ^{+}  & = & A_{++}^{\left[u_{1}\right]d_{2}\left[u_{3}\right]}\left|1_{1}1_{2}1_{3}\right\rangle +A_{-+}^{\left[u_{1}\right]d_{2}\left[u_{3}\right]}\left|1_{1}0_{2}1_{3}\right\rangle \nonumber \\
 & + & \frac{1}{\sqrt{2}}A_{+0}^{\left[u_{1}\right]d_{2}\left[u_{3}\right]}\left(\left|0_{1}1_{2}1_{3}\right\rangle +\left|1_{1}1_{2}0_{3}\right\rangle \right)+\frac{1}{\sqrt{2}}A_{-0}^{\left[u_{1}\right]d_{2}\left[u_{3}\right]}\left(\left|0_{1}0_{2}1_{3}\right\rangle +\left|1_{1}0_{2}0_{3}\right\rangle \right),\;\label{eq:spin-dA}
\end{eqnarray}
where we have expanded the vector diquark with $s_z=+1$ to $|11\rangle$ and $s_z=0$ to $(|01\rangle+|10\rangle)/\sqrt{2}$. $A_{++}^{\cdots}$, $A_{+0}^{\cdots}$, $A_{-+}^{\cdots}$ and $A_{-0}^{\cdots}$ are normalized amplitudes evaluated from Eq.~$\left(\ref{eq:qS-di}\right)$. Comparing Eq.~\eqref{eq:spin-uA} and Eq.~\eqref{eq:spin-dA}, we notice that only the values $A_{\pm+}^{u}$, $A_{\pm0}^{u}$, $A_{\pm+}^{d}$ and $A_{\pm0}^{d}$ are different when the active quark switches from the $u$ quark to the $d$ quark. 

Finally, we sum Eq.~$\left(\ref{eq:spin-qS}\right)$, Eq.~$\left(\ref{eq:spin-uA}\right)$ and Eq.~$\left(\ref{eq:spin-dA}\right)$ together with their respective coefficient $C_{S}$, $C_{V}$ and $C_{VV}$ in Eq.~$\left(\ref{eq:proton-di}\right)$, and obtain the spin state for the valence quarks in the quark-diquark model as,
\begin{eqnarray}
\left|P,+\frac{1}{2}\right\rangle & = & C_{1}\left|0_{1}0_{2}1_{3}\right\rangle +C_{2}\left|0_{1}1_{2}0_{3}\right\rangle +C_{3}\left|1_{1}0_{2}0_{3}\right\rangle +C_{4}\left|0_{1}1_{2}1_{3}\right\rangle \nonumber \\
& +&C_{5}\left|1_{1}0_{2}1_{3}\right\rangle +C_{6}\left|1_{1}1_{2}0_{3}\right\rangle+C_{7}\left|1_{1}1_{2}1_{3}\right\rangle.\;\label{eq:spin-ent}
\end{eqnarray}
With the normalization condition $\langle P,+1/2| P, +1/2\rangle=1$, the values of the seven coefficients are shown on the forth line in Table.~\ref{tab:coefficients}. In addition, we also present the normalized coefficients of the three individual quark-diquark component $|uS^0\rangle$, $|uA^0\rangle$ and $|uA^1\rangle$ following Eq.~$\left(\ref{eq:spin-qS}\right)$, Eq.~$\left(\ref{eq:spin-uA}\right)$ and Eq.~$\left(\ref{eq:spin-dA}\right)$ respectively on the first three lines from Table.~\ref{tab:coefficients}. In the following part of this work, we consider Eq.~\eqref{eq:spin-ent} as the spin state from the quark-diquark model. In the next section, we will study the entanglement structure in the quark-diquark model based on the spin state in Eq.~\eqref{eq:spin-ent}.
\begin{table}
\caption{\label{tab:coefficients} The coefficients of the tripartite spin states of proton based on different models.}
\begin{ruledtabular}
\begin{tabular}{c|c|c|c|c|c|c|c|c}
Model & $\left|000\right\rangle$ & $\left|001\right\rangle$ & $\left|010\right\rangle$ & $\left|011\right\rangle$ & $\left|100\right\rangle$ & $\left|101\right\rangle$ & $\left|110\right\rangle$ & $\left|111\right\rangle$ \tabularnewline
\hline 
$\left|uS^0\right\rangle^{+}$ & $0$ & $0.1752$ & $0.3504$ & $0.3687$ & $0.1752$ & $0.7375$ & $0.3687$ & $0$ \tabularnewline
$\left|uA^0\right\rangle^{+}$ & $0$ & $0.2106$ & $0.4212$ & $0.3789$ & $0.2106$ & $0.6399$ & $0.3789$ & $0.1929$ \tabularnewline
$\left|dA^1\right\rangle^{+}$ & $0$ & $0.3747$ & $0$ & $0.5693$ & $0.3747$ & $0.2226$ & $0.5693$ & $0.1465$ \tabularnewline
quark-diquark~\cite{Maji:2016yqo} & $0$ & $0.2591$ & $0.3407$ & $0.3908$ & $0.2591$ & $0.6546$ & $0.3908$ & $0.1251$ \tabularnewline
BLFQ~\cite{Xu:2021wwj} & $0.0453$ & $0.2998$ & $0.0651$ & $0.3367$ & $0.2998$ & $0.6654$ & $0.3367$ & $0.3800$ \tabularnewline
SU(6) & $0$ & $0$ & $0$ & $0.4082$ & $0$ & $0.8165$ & $0.4082$ & $0$ \tabularnewline
\end{tabular}
\end{ruledtabular}
\end{table}

\section{Spin entanglement signatures from basis light-front quantization\label{sec:Compare-spin-state-BLFQ}}

In this section, we first briefly review the basis light-front quantization (BLFQ) approach and then present the differences in spin entanglement between the light-front wave function from BLFQ and that from the quark-diqaurk model.

Before diving into the details of BLFQ, let us first recall the spin structure of the proton wave function from the naive quark model based on the SU(3) flavor times SU(2) spin symmetry. In this work we denote the wave function from the naive quark model as the SU(6) wave function. When the ordering in the flavor space is fixed as $u_1 d_2 u_3$, the spin part of the SU(6) wave function for the proton state with $J_z =+1/2$ can be written as,
\begin{eqnarray}
\left|P,+\frac{1}{2}\right\rangle _{\mathrm{q}} & = & \frac{1}{\sqrt{6}}\left(2\left|1_{1}0_{2}1_{3}\right\rangle -\left|0_{1}1_{2}1_{3}\right\rangle -\left|1_{1}1_{2}0_{3}\right\rangle \right).\;\label{eq:quark-up}
\end{eqnarray}
By flipping all the qubits we can obtain the $J_{z}=-1/2$ state. In the following study, this SU(6) wave function will serve as a baseline with ``maximal'' spin entanglement due to the presence of significant W-type components, see the discussions below.

The BLFQ is a nonperturbative approach  based on the light-front Hamiltonian formalism~\cite{Vary:2009gt}. In BLFQ, one solves for the structure of relativistic bound states, such as the hadrons, from the light-front Hamiltonian of QCD, see Ref.~\cite{Vary:2025yqo} for a recent review of the application of BLFQ to the proton system. In BLFQ the light-front quantization is adopted~\cite{Brodsky:1997de}. By transforming the Minkowski coordinates $\left(x^{0},x^{1},x^{2},x^{3}\right)$ to light-front coordinates $\left(x^{+},x^{1},x^{2},x^{-}\right)$ $\left(x^{\pm}=x^{0}\pm x^{3}\right)$, the field theory is quantized on the light front, i.e. the hypersurface with equal light-front time $x^{+}$. In the light-front coordinates, the four momentum of a hadron is denoted as $\left(P^{+},P^{-},\boldsymbol{P}_{\perp}\right)$, $P^{\pm}=P^{0}\pm P^{3}$, where $P^{-}$ is the LF Hamiltonian and $P^{+}$ is the momentum in the longitudinal direction. For notational convenience, one often works with the light-front Hamiltonian with the mass squared dimension, which is defined as $H_{\mathrm{LF}}=P_{\mu}P^{\mu}=P^{+}P^{-}-P^2_\perp$. The mass spectroscopy and light-front wave functions of hadrons can be found by solving the eigenvalue equation of the light-front Hamiltonian,
\begin{equation}
H_{\mathrm{LF}}\left|\psi\right\rangle =M^{2}\left|\psi\right\rangle ,\;\label{eq:light-front}
\end{equation}
where $|\psi\rangle$ is the eigenstate of the light-front Hamiltonian $H_{\mathrm{LF}}$ and $M$ is its invariant mass.

In BLFQ, the Hamiltonian eigenequation is solved in a basis approach. At the fixed light-front time $x^{+}=0$, the state of a baryon can be expanded in Fock space as 
\begin{equation}
\left|\Psi\right\rangle =\varphi_{(qqq)}\left|qqq\right\rangle +\varphi_{(qqqg)}\left|qqqg\right\rangle +\varphi_{(qqqq\bar{q})}\left|qqqq\bar{q}\right\rangle \cdots,\ \label{eq:wave-function}
\end{equation}
where $\varphi_{(\cdots)}$ is the light-front wave functions (LFWFs) related to the different Fock sectors $\left|qqq\cdots\right\rangle$. The Fock particles are expressed in terms of single-particle basis. To describe the BLFQ wave functions, a two-dimensional harmonic-oscillator (2D-HO) basis is adopted in the transverse directions, which is labeled by the radial quantum number $n$ and the orbital quantum number $m$. For the degrees of freedom in longitudinal direction, one uses the plane-wave states defined in a one-dimensional box of length $2L$, with periodic boundary conditions for quarks and antiperiodic ones for gluons. These states are labeled by the longitudinal quantum number $k$. The zero modes for gluons are ignored. Together with the light-cone helicity $\lambda$, each parton in a Fock sector is specified by four quantum numbers $\alpha=\{k,n,m,\lambda\}$. Besides, for Fock sectors with more than one color singlet configurations, a color index is introduced to distinguish different color singlet states. In order to make the numerical calculation feasible, two truncation parameters, $\sum_{i}\left(2n_{i}+|m_{i}|+1\right)\le N_{\text{max}}$ and $\sum_{i}k_{i}=K$ are introduced for the transverse and longitudinal directions, respectively. In this work, as an exploratory study, we consider the proton wave function solved in the basis truncated to the leading Fock sector $|udu\rangle$~\cite{Xu:2021wwj}.

The input Hamiltonian consists of an effective one-gluon-exchange interaction as well as a transverse and longitudinal confining interaction motivated by the light-front holographic QCD~\cite{BRODSKY20151}. The one-gluon-exchange interaction is employed to model the spin interaction between the valence quarks. The transverse and longitudinal confining interaction reproduce the three dimensional confinement in the nonrelativistic limit. In this work we consider the light-front wave function calculated with truncation parameters $N_{\rm{max}}$=10 and $K$=16.5. By fitting the proton's mass and electromagnetic form factors, one determines the input parameters: $m_{{\rm {q/k}}}=0.3\,\mathrm{GeV}$ and $m_{{\rm {q/g}}}=0.2\,\mathrm{GeV}$ denoting the mass of quarks in the kinetic terms and the one-gluon-exchange interaction terms respectively, and the confining strength $\kappa=0.34\,\mathrm{GeV}$, and the strong coupling constant $\alpha_{s}=1.1\pm0.1$, see Ref.~\cite{Xu:2021wwj} for the details on the input parameters. Similar to the wave function  from the quark-diquark model, we calculate the coefficients for each spin configuration from the BLFQ wave function for the proton and list the result in Table.~\ref{tab:coefficients}. In order to examine how the input parameters (such as the mass and the coupling constant) influence the spin entanglement among the valence quarks, we vary their values and compare the entanglement patterns of the resulting LFWFs in Sec.~\ref{sec:Spin-ent-para}.

\begin{figure}
\includegraphics[scale=0.8]{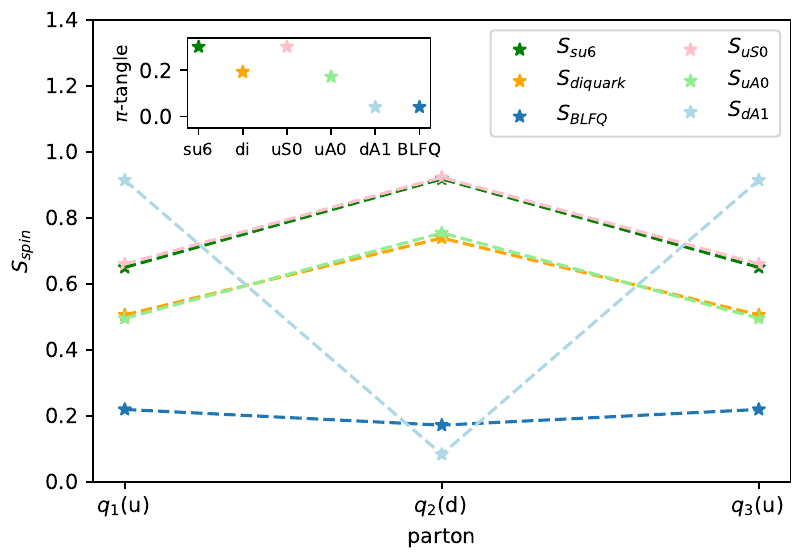}
\caption{\label{fig:ent-compare-models} \footnotesize The comparison of bipartite entanglement and tripartite entanglement between the spin states from three different models: the quark-diquark model~\cite{Maji:2016yqo}, the BLFQ model~\cite{Xu:2021wwj}, and the $\mathrm{SU\left(6\right)}$ quark model. The three spin sectors from the quark-diquark model are included as well (dots and dashed lines in light colors). The main plot shows the entanglement entropy of one parton with the remaining two partons of the proton states from different models, and the inset shows the $\pi$-tangle among the three partons.}
\end{figure}

To study the entanglement structures of various spin states in Table.~\ref{tab:coefficients}, we use the entanglement entropy introduced in Eq.~$\left(\ref{eq:ent-entropy}\right)$ to measure their bipartite entanglement and the $\pi-$tangle in Eq.~$\left(\ref{eq:pi-tangle}\right)$ to measure their tripartite entanglement. Since the triangle measure $F_{123}$ in Eq.~\eqref{eq:F123} produces qualitatively similar results, here we only present the $\pi-$tangle as an example. The results are presented in Fig.~\ref{fig:ent-compare-models}. We find that both bipartite and tripartite entanglement of the spin states from the quark-diquark model is significantly larger than that from BLFQ, and they are also closer to the ``maximal'' entanglement found in the naive SU(6) quark model. This can be understood as follows: when the diquark cluster is resolved as a two-quark cluster state, additional entanglement arises from the strong correlations in the two-quark state, whereas in the BLFQ framework the three quarks are treated as independent constituents. We note that in principle the LFWF for the proton includes not only valence quarks but also gluons and sea quarks. The LFWF in the truncated bases may only captures a part of the entanglement information in the full proton state. Based on our previous study~\cite{Qian:2024fqf}, once one dynamical gluon is included in the basis ($|qqq\rangle$+$|qqqg\rangle$) the entanglement entropies of the spin of the $\{u,d,u\}$ quarks become $\{0.414, 0.483, 0.414\}$, which are significantly closer to the quark-diquark model values, $\{0.506, 0.739, 0.506\}$, compared to the corresponding results in the $\left|qqq\right\rangle$ sector, $\{0.219, 0.172, 0.219\}$.

In Fig.~\ref{fig:ent-compare-models}, we also depict the bipartite and tripartite entanglement of the three types of quark-diquark spin sectors $\left|uS^0\right\rangle$, $\left|uA^0\right\rangle$ and $\left|dA^1\right\rangle$ for reference. The exchange (anti-)symmetry between the $u$ and $d$ quarks injects substantial entanglement into the $\left|uS^0\right\rangle$ and $\left|uA^0\right\rangle$ sectors. From the coefficients in Table.~\ref{tab:coefficients}, $\left|uS^0\right\rangle$ can be identified as a superposition of the $\mathrm{SU\left(6\right)}$ spin-up state and the spin-down state, thus it has the largest entanglement for all the models compared in this work except for the $\mathrm{SU\left(6\right)}$ state.

In terms of the classification of tripartite entangled states discussed in Sec.~\ref{sec:methodology},  distinct entanglement signatures in the spin configurations from the quark-diquark model and from BLFQ can be identified. According to Table~\ref{tab:coefficients}, the quark-diquark and BLFQ spin configurations differ most prominently in the $\left|000\right\rangle$ and $\left|010\right\rangle$ components: the coefficient of $\left|000\right\rangle$ vanishes in the quark-diquark state but is finite in BLFQ, whereas the coefficient of $\left|010\right\rangle$ is much larger in the quark-diquark state than in BLFQ. According to the tripartite qubit state classification introduced in Sec.~\ref{sec:methodology}, the $\left|000\right\rangle$ component is associated with GHZ-type entanglement and the $\left|010\right\rangle$ component is associated with W-type entanglement. In the quark-diquark model, the tripartite state is described as either a Bell-type biseparable state or a W-type state which can be generated from the Bell-type state through permutation. Its W-type state consists of both the $2-$excitation Dicke components $\{|011\rangle, |101\rangle, |110\rangle\}$ and the $1-$excitation Dicke components $\{|001\rangle, |010\rangle, |100\rangle\}$. The GHZ-type entanglement is not expected in the quark-diquark model, and the fully separable states are suppressed. On the other hand, the BLFQ spin state consists of W-type entanglement from the $2$-excitation Dicke components $\left\{ \left|011\right\rangle ,\left|101\right\rangle ,\left|110\right\rangle \right\}$, Bell-type biseparable entanglement from the spin components $\left\{ \left|001\right\rangle ,\left|100\right\rangle\right\}$, and a small amount of GHZ-type entanglement from the components $\left\{ \left|000\right\rangle ,\left|111\right\rangle\right\}$. Although the GHZ-type entanglement is genuinely tripartite compared to the W-type entanglement and typically yields a larger $\pi-$tangle, in the quark-diquark spin state, the W-type component takes a much larger percentage than the combined W- and GHZ-type component in the BLFQ spin state. Therefore, the quark-diquark spin state produces much larger the $\pi-$tangle compared to that for the BLFQ.

\section{Entanglement structures with different parameters in BLFQ\label{sec:Spin-ent-para}}

In this section, we investigate the sensitivity of the entanglement structure to the input parameters in the light-front Hamiltonian in BLFQ. In the previous section, we have adopted the parameter set $\{N_{\rm{max}},K,m_{{\rm {q/k}}},\alpha_{s}\}=\{10,16.5,0.3,1.1\}$, which is considered an optimal parameter set to reproduce the proton's electromagnetic properties~\cite{Xu:2021wwj}. In this section, we vary the value of the quark mass and the strong coupling constant and compare the entanglement structure in the resulting LFWFs with that from the quark-diquark model. As we have found that the entanglement structure depends sensitively on the spin configurations in the proton wave functions, a natural question arises that, by varying the input parameters in the light-front Hamiltonian, how much the spin entanglement structure in the resulting LFWF will change accordingly.

\begin{figure}
\includegraphics[scale=0.55]{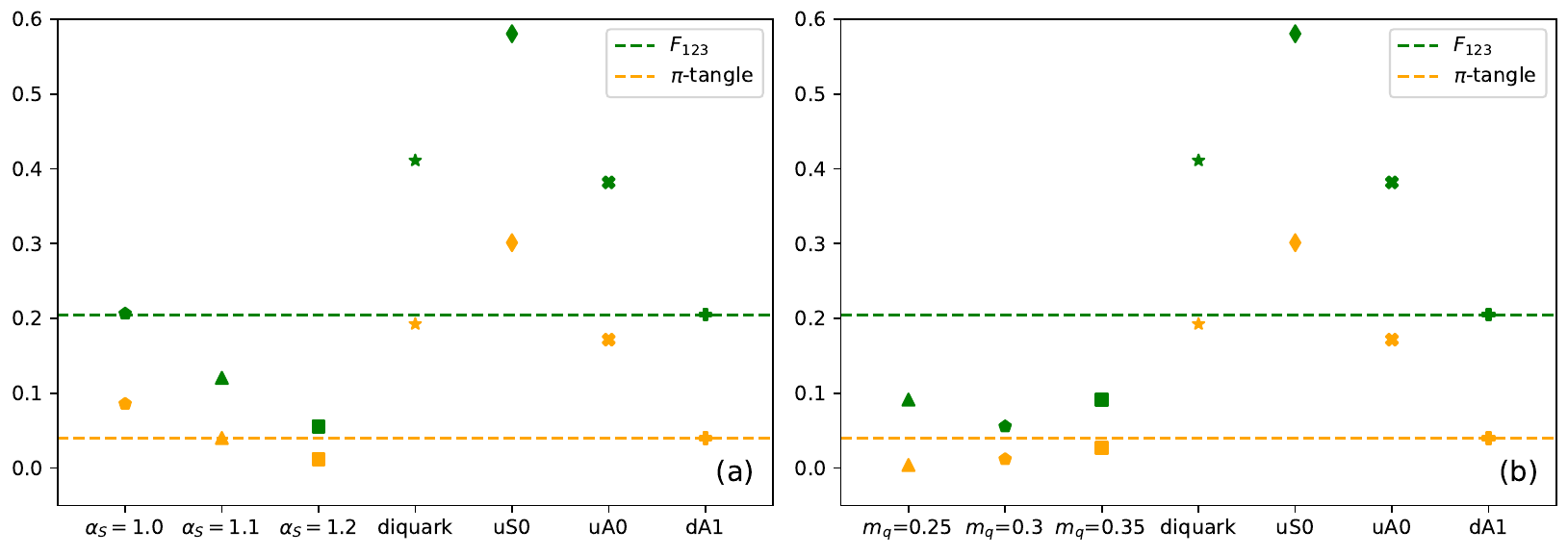}
\caption{\label{fig:ent-tri-6} \footnotesize The comparison of the tripartite entanglement between the BLFQ spin states~\cite{Xu:2021wwj} with different parameters and those from the quark-diquark model~\cite{Maji:2016yqo}. The green dots are the values of the triangle measure $F_{123}$ and the orange dots are the values of the $\pi-$tangle. The dashed lines indicate the $F_{123}$ and $\pi$-tangle values of the $d-uu$ quark-diquark state $\left|dA^1\right\rangle$, see text for details. The subplots: (a) Tripartite entanglement with the strong coupling constant $\alpha_{s}=1.0$, $1.1$, $1.2$ and quark mass $m_\mathrm{q/k}=0.3\,\rm{GeV}$. (b) Tripartite entanglement with the quark mass $m_{\mathrm{q/k}}=0.25\,\mathrm{GeV}$, $0.3\,\mathrm{GeV}$, $0.35\,\mathrm{GeV}$ and the strong coupling constant $\alpha_s=1.2$.}
\end{figure}

\begin{figure}
\includegraphics[scale=0.55]{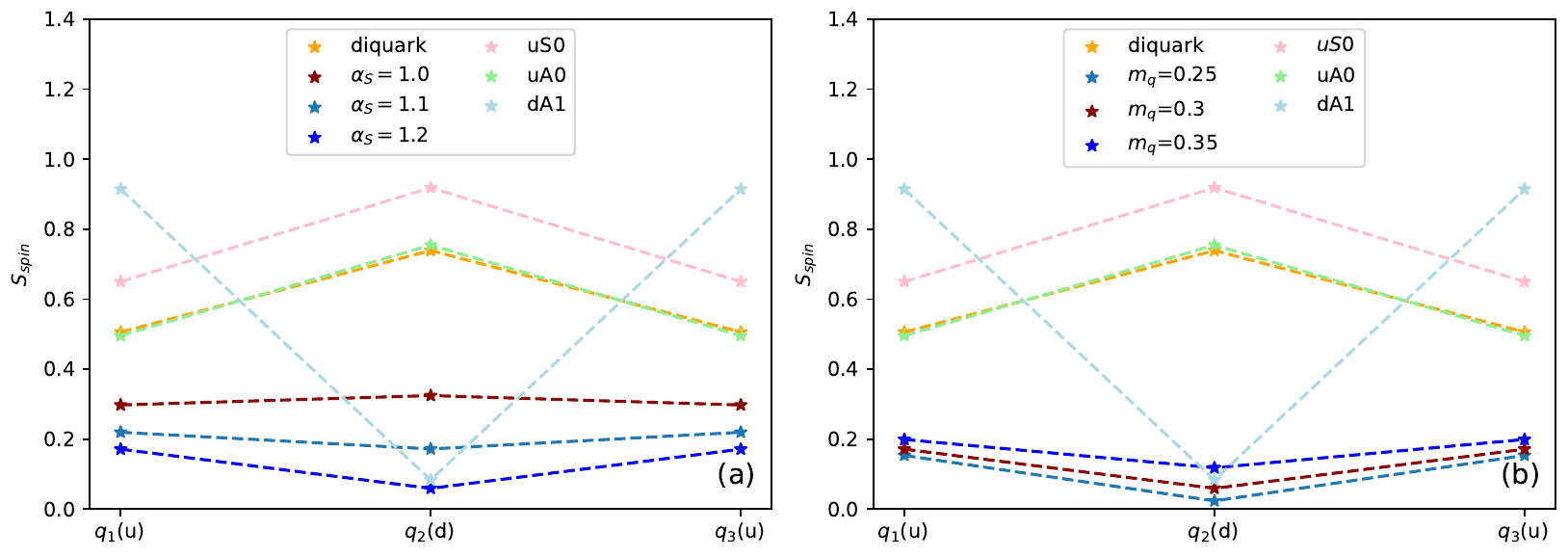}
\caption{\label{fig:ent-bi-6} \footnotesize The comparison of the bipartite entanglement entropy between the BLFQ spin states~\cite{Xu:2021wwj} with different parameters and those from the quark-diquark model~\cite{Maji:2016yqo}. The dots and dashed lines indicate the entanglement entropy of the quark-diquark spin state and the three individual spin sectors, see text for details. The subplots: (a) Entanglement entropy with strong coupling constant $\alpha_{s}=1.0$, $1.1$, $1.2$ and the quark mass $m_\mathrm{q/k}=0.3\,\rm{GeV}$. (b) Entanglement entropy with the quark mass $m_{\mathrm{q/k}}=0.25\,\mathrm{GeV}$, $0.3\,\mathrm{GeV}$, $0.35\,\mathrm{GeV}$ and strong coupling constant $\alpha_s=1.2$.}
\end{figure}

We present the resulting coefficients of the BLFQ spin states with different values of the quark mass in the kinetic energy  $m_{\mathrm{q/k}}$ and the strong coupling constant $\alpha_s$ in Table.~\ref{tab:coefficients-blfq}. For simplicity, we keep the quark mass in the one-gluon-exchange interaction fixed at $m_{\mathrm{q/g}}=0.2\;\rm{GeV}$. We show the tripartite entanglement characterized by the $\pi$-tangle and the triangle $F_{123}$ in Fig.~\ref{fig:ent-tri-6} and the bipartite entanglement characterized by the entanglement entropy in Fig.~\ref{fig:ent-bi-6} for different values of $\{m_{\mathrm{q/k}},\alpha_s\}$. In Figs.~\ref{fig:ent-tri-6} and \ref{fig:ent-bi-6} we also present the results of the quark-diquark spin state and the three sectors for comparison. From Figs.~\ref{fig:ent-tri-6} and \ref{fig:ent-bi-6}, we observe that the variations of $\{m_{\mathrm{q/k}}, \alpha_s\}$ do not significantly influence the qualitative structure of entanglement in the BLFQ spin states. Thus, the difference in spin entanglement between the quark-diquark model and the BLFQ approach can be attributed to the fundamentally different treatment of the proton's internal degrees of freedom rather than the choice of the input parameters. 

\begin{table}
\caption{\label{tab:coefficients-blfq} The coefficients of the BLFQ spin states of a proton with truncation parameters $N_{\rm{max}}=10$ and $K=16.5$, where the quark mass in the kinetic energy and coupling constants are varied, see text for details.}
\begin{ruledtabular}
\begin{tabular}{c|c|c|c|c|c|c|c|c|c}
$m_{\mathrm{q/k}}$ & $\alpha_{s}$ & $\left|000\right\rangle$ & $\left|001\right\rangle$ & $\left|010\right\rangle$ & $\left|011\right\rangle$ & $\left|100\right\rangle$ & $\left|101\right\rangle$ & $\left|110\right\rangle$ & $\left|111\right\rangle$ \tabularnewline
\hline 
$0.3$ & $1.0$ & $0.0354$ & $0.2618$ & $0.0601$ & $0.3554$ & $0.2618$ & $0.7062$ & $0.3554$ & $0.3266$ \tabularnewline
$0.3$ & $1.1$ & $0.0453$ & $0.2998$ & $0.0650$ & $0.3367$ & $0.2998$ & $0.6654$ & $0.3367$ & $0.3800$ \tabularnewline
$0.3$ & $1.2$ & $0.0539$ & $0.3309$ & $0.0679$ & $0.3178$ & $0.3308$ & $0.6234$ & $0.3178$ & $0.4279$ \tabularnewline
$0.25$ & $1.2$ & $0.0569$ & $0.3406$ & $0.0729$ & $0.3092$ & $0.3406$ & $0.6007$ & $0.3092$ & $0.4554$ \tabularnewline
$0.35$ & $1.2$ & $0.0498$ & $0.3163$ & $0.0638$ & $0.3284$ &  $0.3163$ & $0.6480$ & $0.3284$ & $0.3971$ \tabularnewline
\end{tabular}
\end{ruledtabular}
\end{table}

In addition, we notice that as the strong coupling $\alpha_s$ increases and the quark mass $m_{\mathrm{q/k}}$ decreases, the entanglement structure of the BLFQ spin state approaches that of the $d-uu$ quark-diquark configuration, namely $|dA^1\rangle$. The state $|dA^1\rangle$, formulated in Eq.~\eqref{eq:spin-dA}, describes a quark-diquark configuration where the active $d$ quark is approximately separable from the remaining diquark subsystem, where the $uu$ pair is almost Bell-entangled. As shown in Figs.~\ref{fig:ent-tri-6} and \ref{fig:ent-bi-6}, the genuine tripartite entanglement of the BLFQ spin states among the three quarks $udu$ gradually approaches zero as $\alpha_s$ and smaller $m_{\mathrm{q/k}}$, whereas the entanglement between the two $u$ quarks remains nonvanishing. Therefore, the $|udu\rangle$ state tends to approach the factorized form $\left|d\right\rangle\otimes\left|uu\right\rangle$ in this limit.

The origin of this trend can be understood from the coefficients in Table~\ref{tab:coefficients-blfq}. When $\alpha_s$ increases from $1.0$ to $1.2$ at fixed $m_{\mathrm{q/k}}=0.3$, the coefficient of $|101\rangle$ decreases from $0.7062$ to $0.6234$. This component is associated with the W-type entanglement between $\{|101\rangle,|011\rangle,|110\rangle\}$ in the framework of tripartite state classification, thus its reduction indicates that the W-type entanglement degrades as the coupling becomes stronger. At the same time, the coefficients of $|001\rangle$ and $|100\rangle$ increase from $0.2618$ to $0.3309$. These two components are associated with Bell-type entanglement between the two $u$ quarks, and their enhancement shows that the bipartite $uu$ correlation becomes more prominent. A similar trend is observed when varying the quark mass at fixed $\alpha_s=1.2$: as $m_{\mathrm{q/k}}$ decreases from $0.35~\mathrm{GeV}$ to $0.25~\mathrm{GeV}$, the coefficient of $|101\rangle$ decreases from $0.6480$ to $0.6007$, while those of $|001\rangle$ and $|100\rangle$ increase from $0.3163$ to $0.3406$. Therefore, a larger coupling $\alpha_s$ and a smaller quark mass $m_{\mathrm{q/k}}$ act in the same direction: they suppress the W-type contribution and enhance the Bell-type entanglement, thereby driving the BLFQ spin state towards the $|dA^1\rangle$ configuration. 

Nevertheless, we note that even in the large coupling and small quark mass limit the BLFQ spin state remains quantitatively distinct from the ideal quark-diquark configuration. Specifically, as $\alpha_s$ increases from $1.0$ to $1.2$ with the quark mass $m_{\mathrm{q/k}}$ fixed at $0.3\,\mathrm{GeV}$, the coefficient of $|111\rangle$ rises from $0.3266$ to $0.4279$; similarly, as $m_{\mathrm{q/k}}$ decreases from $0.35~\mathrm{GeV}$ to $0.25~\mathrm{GeV}$ with the coupling constant $\alpha_s$ fixed at $1.2$, it increases from $0.3971$ to $0.4554$. This means that, although the Bell-type components $|001\rangle$ and $|100\rangle$ become more important in the limit of large coupling constant or small quark mass, the sizable $|111\rangle$ contribution weakens the Bell-type entanglement in the $uu$ subsystem.

\section{Conclusions and outlook\label{sec:Conclusions-and-outlook}}

In this work, we investigate the spin entanglement structure of the proton computed from the BLFQ approach in the basis truncated to the $|qqq\rangle$ Fock sector, and compare it with that predicted by a light-front quark-diquark model. We analyze the state classification of the tripartite spin state components, and calculate the bipartite and tripartite entanglement quantities. Based on these results, we conclude that both the bipartite and tripartite entanglement in the BLFQ spin state are significantly smaller than those in the quark-diquark model. This difference can be traced back to the distinct physical pictures adopted in the two frameworks: in BLFQ the proton is described in terms of three independent constituent quarks, whereas in the quark-diquark model it is treated as an active quark coupled to a correlated diquark cluster. 

To make a fair comparison, we resolve the diquark cluster as a correlated two-quark cluster and expand the corresponding spin wave function in the three-qubit basis. We quantify bipartite entanglement by the entanglement entropy and tripartite entanglement by the $\pi-$tangle. We demonstrate that the BLFQ spin state exhibits only weak entanglement compared to both the quark-diquark state and the naive SU(6) quark model. To further discriminate the entanglement patterns, we analyze the coefficients of the different spin states and observe that, owing to the exchange symmetry of the diquark cluster, the W-type entanglement in the quark-diquark state is much stronger than that in the BLFQ spin state. In addition, the relatively large weight of the $\left|111\right\rangle$ component suppresses both bipartite and tripartite entanglement in the BLFQ. However, since the BLFQ proton wave function could in principle contain not only the $\left|qqq\right\rangle$ but also higher Fock sectors involving gluons and sea quarks, we expect the entanglement to increase when higher Fock sectors are included.

Moreover, we vary several parameters in the input light-front Hamiltonian of BLFQ to examine their impact on the entanglement of resulting proton light-front wave functions. We conclude that different parameter choices do not significantly modify the basic bipartite and tripartite entanglement structure of the BLFQ spin states. Nevertheless, when the coupling is strong and the quark mass is small, the BLFQ spin state develops a configuration similar to the $\left|dA^1\right\rangle$ component of quark-diquark model, where the $d$ quark behaves as an active quark and the two $u$ quarks form a diquark cluster. However, the sizable presence of the $\left|111\right\rangle$ components weakens the Bell-type entanglement in the BLFQ spin states, making them qualitatively different from the $\left|dA^1\right\rangle$ configuration.

As a future direction, this work motivates us to perform quantum state tomography for partons inside hadrons using the tools from the quantum information theory. This topic has recently attracted increasing attention in high-energy physics~\cite{Afik:2020onf, Aguilar-Saavedra:2024vpd,Wu:2024asu,Altomonte:2024upf,Afik:2025ejh,Wu:2025dds}. If the information about partonic entanglement can be experimentally accessed in the future, it could serve as a powerful avenue for connecting partonic structure with the fundamental interactions. The BLFQ approach is particularly suited for this purpose, since all the quantum correlation information between partons are encoded in the many-body light-front wave functions. We envision that studying quantum-information-theoretic correlations, such as the entanglement between its spin, momentum, or color degrees of freedom, using the LFWFs from BLFQ can provide novel perspectives on hadrons' nonperturbative structure in terms of QCD first principles.

\begin{acknowledgments}
We thank Yun-Heng Ma for support with the \emph{Qton} software used for the entanglement entropy calculations. C. Q. is supported by the National Key Research and Development Program of China (2025YFE0217200) and the National Natural Science Foundation of China (NSFC) with Grant No. 12305010. Y. G. Y. is supported by Postdoctoral Program of Gansu Province. X. Z. is supported by National Natural Science Foundation of China, Grant No. 12375143, new faculty startup funding by the Institute of Modern Physics, Chinese Academy of Sciences, by Key Research Program of Frontier Sciences, Chinese Academy of Sciences, Grant No. ZDBS-LY-7020, by the Foundation for Key Talents of Gansu Province, by the Central Funds Guiding the Local Science and Technology Development of Gansu Province, Grant No. 22ZY1QA006, by Gansu International Collaboration and Talents Recruitment Base of Particle Physics (2023-2027), by International Partnership Program of the Chinese Academy of Sciences, Grant No. 016GJHZ2022103FN, by National Natural Science Foundation of China, Grant No. 12375143, by National Key R\&D Program of China, Grant No. 2023YFA1606903 and by the Strategic Priority Research Program of the Chinese Academy of Sciences, Grant No. XDB34000000.
\end{acknowledgments}

\bibliography{ref-tri}
\end{document}